# The Future Orchid Diversity of Great Britain and Ireland using an SDM Approach

Sofoklis Mouratidis, Konstantinos Kougioumoutzis, Martha Charitonidou, John M. Halley

[1] Department of Biological Applications and Technology, University of Ioannina, 45110 Ioannina, Greece
[2] Laboratory of Botany, Department of Biology, University of Patras, GR-26504 Patras, Greece

* These authors contributed equally
** Corresponding author: Department of Biological Applications and Technology, University of Ioannina, 45110 Ioannina, Greece. Telephone: +30 26510-07337, email: jhalley@uoi.gr

**Abstract**
In this paper we use Species Distribution Models (SDMs) to forecast the future diversity and distribution of orchids in Great Britain and Ireland under scenarios of climate and land-use change. The study analyzes occurrence data for native orchid taxa in the BSBI database at a fine spatial resolution (1 km^2, monads) and incorporates multiple environmental variables including climate, land use, topography, and soil. These SDMs project significant losses in orchid species richness by 2050 and 2070, especially under severe climate and land-use scenarios, with declines expected across most species and regions, including Ireland where historical data previously indicated gains. The models reveal vulnerable species likely to face extinction by 2070, emphasizing the impact of both climate warming and habitat modifications. This approach differs from previous trend-based analyses by integrating future projections, high-resolution spatial data, and dynamic land-use scenarios, thereby providing higher-resolution estimates of orchid range contractions and diversity losses. While current observed orchid trends show some regional increases, particularly in Ireland, the SDM forecasts indicate substantial future risks. The study also discusses uncertainties due to niche truncation from geographic data limits and highlights the need for broader-scale modeling for more robust predictions. Overall, the paper anticipates conservation challenges for orchid biodiversity in response to ongoing environmental changes.

**Keywords**
migration; occupancy trends; Orchidaceae; land use change; climate change; orchid distribution; range shift; diversity hotspots; species distribution models (SDMs)

**Declarations**
The research project was supported by the Hellenic Foundation for Research and Innovation (H.F.R.I.) under the "2nd Call for H.F.R.I. Research Projects to support Faculty Members & Researchers" (Project Number: 3972).

**Conflicts of interest/Competing interests (include appropriate disclosures)**
Authors have no competing interests

**Availability of data and material (data transparency)**
The data analyzed in this study were obtained from the Botanical Society of Britain and Ireland (BSBI) Distribution Database (DDb: https://database.bsbi.org). Data from the BSBI are subject to copyright and database rights described in the Data Access Policy (https://docs.bsbi.org/using-our-data/data-access-policy). Requests to access these datasets should be directed to BSBI Distribution Database-Data Team, dataaccess@bsbi.org.

**Code availability (software application or custom code)**
The code used in this study is available at: https://github.com/EcoLabBET/ExOrChiST.git.

**Ethics approval (include appropriate approvals or waivers)**
Not applicable

**Consent to participate (include appropriate statements)**
Not applicable

**Consent for publication (include appropriate statements)**
All authors declare that neither the article nor portions of it have been previously published elsewhere; all authors declare that the manuscript is not under consideration for publication in another journal; all authors consent to the publication of the manuscript in Biodiversity and Conservation, should the article be accepted by the Editor-in-chief upon completion of the refereeing process

# 1 Introduction

Fifty-seven species of orchid are presently considered native to Britain and Ireland, although this is subject to frequent taxonomic changes and immigration events (Bateman, 2022). Great Britain is one of the most highly populated and developed regions of Europe, especially in areas where orchid diversity is highest in southern and south-eastern England. Since the 1940s, much of the British countryside has been heavily modified by agricultural changes, mainly the intensification of production, which led to the large-scale loss of semi-natural habitat and increasing nutrient inputs (Burns et al., 2016; Montràs-Janer et al., 2024; Taheri et al., 2021). Research using digitised historical maps has revealed a loss of over half (> 3000 km²) of the semi-natural grasslands in Great Britain over the last 75 years (Suggitt et al., 2023).

Because one of the major orchid habitats is semi-natural grassland, land-use change is regarded as a major threat to orchids and a major concern in orchid conservation. Concerns about the populations of orchids are widespread and several scientific studies have argued that orchids are "in decline" in Britain and Europe (Damgaard et al., 2020; Kull & Hutchings, 2006; Trudgill, 2022; Vogt-Schilb et al., 2015). However, definitive answers remain elusive.

Occurrence records for the orchid flora of Great Britain and Ireland are available in the BSBI's database of distribution for British and Irish vascular plants from 1930 to 2019 (Walker et al., 2023). In a recent study (Charitonidou et al., 2025) for both Great Britain and Ireland over the 90-year period covered by the BSBI atlases, changes in the distribution of 51 of these orchid species were analysed at the hectad (10km × 10km) level. The study focused on trends of orchid occupancy, centroid movement and the redistribution of orchid biodiversity in Great Britain and Ireland, using corrections for uneven recording and for temporal autocorrelation. The study found a dominant pattern of decline in orchid biodiversity, with the number of high-diversity hotspots ($SR \geq 16$ species) decreasing (from 107 to 41). Also, eight of the 13 significant trends (at the 5% level) were decreasing in Great Britain. No convincing evidence of climate-related shift could be seen – while orchids did seem to be moving North in Great Britain, the trend is clearly south-westwards in Ireland. The results for Great Britain broadly support the conclusions of others that there is a decline in orchid occurrence in the UK. However, the results for Ireland go against this pattern, with four out of five significant trends positive and six high-diversity hotspots appearing in recent years. The inclusion of temporal autocorrelation ("natural trends and cycles") clearly demonstrates that fewer changes can be taken to be significant departures from natural variability. Failure to detect significant climatic signal is in agreement with recent publications investigating range shifts in Britain, using atlases of the BSBI, did not find evidence for clear patterns of change in plant distributions (Groom, 2013; Montràs-Janer et al., 2024). Nevertheless, the results leave a number of questions. Firstly, could the orchid increases in Ireland, as well as the southwest centroid drift, be an artifact of better recording in recent years? Secondly, might the rather nuanced declines in occupancy in Great Britain be an artifact of the coarse scale (hectads) employed and might it be more visible on a finer scale, such as monads. While land use change is currently regarded as a major threat to orchid biodiversity in Europe, the signal of climate change is not yet visible. However, climate change is projected to become a dominant driver in the future (Moreira et al., 2023; Pimm, 2008).

In this study, we proceed to use the BSBI database to forecast the future, through the use of species distribution models (SDMs) closely following an approach pioneered for the flora of Evvia,

Greece (Kougioumoutzis et al., 2025). The SDMs here incorporate both future land-use and climate-change scenarios. In addition, in this study the spatial resolution will be finer (monad level, 1km×1km) rather than the coarser hectad resolution. By applying SDMs to forecast future distributions, we can anticipate the conservation problems that will arise, and for which species. We will also have a clearer idea of when the climatic effects are likely to become more significant. Also, incorporating dynamic use/land cover change (LULC) data along with the climate change scenarios, we can improve our vulnerability assessments for the orchid flora. Our approach to modelling the entire orchid flora thus represents a significant advance in understanding the future of orchid biodiversity of the British Isles. By integrating LULC data alongside climate projections into our SDMs, this study lays the groundwork for addressing better the relative importance of CC and LULC on future and present orchid biodiversity patterns. Our analysis can also deepen our understanding of how these taxa may adapt to or be impacted by forthcoming ecosystem alterations.

Specifically, our research objectives are:

1. To evaluate, using SDMs, species-specific responses to expected climate and land-use change for time-periods centred on 2050 and 2080 (Kougioumoutzis et al., 2025).
2. To extend our trend analysis of orchid occupancy into the future and assess expected increases or decreases
3. Assessment of taxon-specific extinction threats to orchids at these future time-periods
4. Prediction of behaviour of orchid hotspots in the future

## 2 Methods

### 2.1 Species Occurrence Data

Our study concentrates on the orchid taxa found in Great Britain (including the Isle of Man; GB) and Ireland (IR), treated as a combined geographical unit. Occurrence data were sourced from the BSBI's comprehensive database for British and Irish vascular plants distribution records (Stroh et al., 2023). This database holds records for 57 orchid species documented in Britain and Ireland (Bateman, 2022) (see Appendix A). We chose species based on having a minimum of 10 occurrences (ten occupied monads), as well as ecological and taxonomic criteria (Charitonidou et al., 2025), resulting in a subset of 45 species included in the analyses for GB and IR (Appendix A). Occurrence data were downloaded from the BSBI database (https://database.bsbi.org/ - last accessed February 2024) and subsequently converted into presence/absence data within monads for each defined time period. Our taxonomic framework follows Bateman (2022).

Alpha hulls, computed using the 'EOO.computing' function within the 'ConR' 1.3.3 package (Dauby et al., 2017), were applied to outline the distributional ranges of the target species across GB and IR. This technique excludes gaps in ranges and performs well with habitats of irregular shape or uneven sampling coverage (Burgman & Fox, 2003). We applied the 'sdm_extract' function from the 'flexsdm' R package to discard occurrences with NA values in any abiotic variables (refer to Section 2.2), which were incorporated in the analysis. Occurrence data were further refined using the 'clean_coordinates' function from the 'CoordinateCleaner' 2.0.18 package (Zizka et al., 2019) to remove potentially erroneous records. Duplicates were eliminated employing the 'elimCellDups' function from the 'enmSdm' 0.5.3.3 package (Smith, 2020). Spatial thinning to one record per 1 km² was performed using the 'thin' function from the 'spThin' 0.1.0 package (Aiello-Lammens et al., 2015) to align with the spatial resolution of predictor variables. Our cleaning and spatial thinning followed established protocols and Species Distribution Model (SDM) guidelines (Araújo et al., 2019; Soley-Guardia et al., 2024), assuring data quality and spatial consistency. Consistent with Kougioumoutzis et al. (2025), analyses were limited to taxa with at least ten occurrences, meeting the requirements of the Ensemble of Small Models (ESMs) framework (see Section 2.3), which can yield reliable distribution models even with as few as two occurrences per taxon (Erickson & Smith, 2023). The final dataset comprised nearly 62,000 occurrence records covering 45 taxa (Appendix A).

**2.2 Environmental and Land-Use Change Data**

We developed a high-resolution (1 km) monthly climate dataset for the year 2015, which includes 19 bioclimatic variables from WorldClim (Fick & Hijmans, 2017) and 16 additional environmental variables (Title & Bemmels, 2018). This resolution matches that of the occurrence records. Using 2015 as the baseline year ensured compatibility among all abiotic variables. The spatial resolution was set at 1 km, and altitude data were integrated using the 'elevation_30s' function from the geodata 0.6-2 R package. The dataset also incorporated processing performed with ClimateEU v4.63 and the R packages "dismo" 1.1.4 (Hijmans et al., 2017) and "envirem" 2.2 (Title & Bemmels, 2018), following protocols described in Kougioumoutzis et al. (2025). Specifically, the altitudinal data helped derive monthly climate data via ClimateEU v4.63. From this, we generated 37 climatic variables using the 'biovars', 'ETsolradRasters', and 'generateEnvirem' functions from the "dismo" and "envirem" packages.

Dynamic land use data at 1 km resolution were sourced from Chen et al. (2022), providing the finest available global land use projections. Their original 20 land use categories were transformed into separate binary predictor variables for analysis. Soil variables were incorporated from SoilGrids (Hengl et al., 2017) at the same spatial resolution as other environmental metrics, following established SDM guidelines (Araújo et al., 2019; Soley-Guardia et al., 2024). Five key topographical metrics—aspect, heat load index, slope, topographic position index, and terrain ruggedness index—were calculated using altitude data and functions from the R packages 'terra' 1.7.46 (Hijmans, 2023) and "spatialEco" 1.2-0 (Evans, 2019). Rainfall soil erosivity data were obtained with the 'soil_world' function from the 'geodata' R package (Hijmans et al., 2024) and from Panagos et al. (2022), maintaining spatial resolution consistency across variables.

Climate projections were generated for two future periods: the 2050s (2041–2070) and the 2080s (2071–2100). Rather than use WorldClim's CMIP6 future climate projections, which base their historical data on 1970–2000 and would not align with our 2015 baseline for land use, we selected two CMIP5 global circulation models available via ClimateEU, following the selection criteria in Kougioumoutzis et al. (2025). These included CCSM4, HadGEM2, and an ensemble of 15 GCMs to better encompass uncertainty from different climate model structures. Each model incorporated two IPCC emission scenarios: the moderate RCP4.5 and the more extreme RCP8.5.

Corresponding future land use and land cover (LULC) projections from Chen et al. (2022) covered two Shared Socioeconomic Pathways (SSPs): SSP1-RCP2.6, SSP3-RCP7.0, and SSP5-RCP8.5 (Cao et al., 2022). The original 20 land use categories were consolidated into six broader classes—'forests', 'shrubs', 'grasslands', 'barren', 'crops', and 'urban'—using the 'terra' package, applied consistently across baseline and projected LULC datasets.

Among the environmental variables, topographic and soil metrics were considered static over time, while bioclimatic, soil erosivity, and LULC variables were treated as dynamic. These predictors represent critical ecophysiological factors influencing plant survival, encompassing temperature, water availability, and light—key drivers of plant species distributions (Mod et al., 2016). To remove highly collinear variables, we employed Spearman rank correlations (<0.7) and variance inflation factors (<5) using the 'collinear' 1.1.1 R package for precise collinearity diagnostics.

**2.3 Species Distribution Models**

Our species distribution modeling approach primarily follows Kougioumoutzis et al. (2025) with some adjustments; further details are in Appendix 2 (S.I. The ODMAP Protocol). The focus is on orchids in Great Britain and Ireland, many of which are specialized with narrow ecological niches and limited occurrence data. To handle taxa with sparse records, we applied the Ensemble of Small Models (ESMs) framework, which effectively models species distributions even for taxa with as few as two occurrence points (Erickson & Smith, 2023). Our analyses included taxa with occurrence-to-predictor ratios under 10:1, adhering to ESM modeling guidelines (Breiner et al., 2018). Modeling used the Random Forest algorithm with 1000 trees ('ntree'=1000) and the 'ecospat.ESM.Modeling' and 'ecospat.ESM.EnsembleModeling' functions from the "ecospat" 3.1 R package (Broennimann et al., 2021). The ESM method's reliance on bivariate models reduces simultaneous predictor use, helping mitigate issues associated with low occurrence-to-predictor ratios compared to full multi-predictor models.

Taxa were divided into two groups for pseudo-absence generation: species with ≥10 occurrences and those with 5–9 occurrences (Jimenez-Valverde, 2021). The first group's pseudo-absences were generated with the 'sample_pseudoabs' function from 'flexsdm' 1.3.0 using the 'geo_env_km_const' method (Velazco et al., 2022). The second group utilized random pseudo-absences per protocols for rare-specialist taxa (Inman et al., 2021). For taxa with ≥20 occurrences, we partitioned data via 'bm_CrossValidation' in 'biomod' 4.2.4 with the 'block' strategy (Thuiller et al., 2016; Kougioumoutzis et al., 2025). Taxa with 5–19 occurrences were cross-validated randomly with 10 repetitions and a 75% train / 25% test split.

Models achieving a minimum TSS of 0.4 (Engler et al., 2011) in 'ecospat.ESM.EnsembleModeling' were retained to identify suitable habitats, consistent with established thresholds (Franklin, 2010). For robust reliability, we kept models meeting pooled TSS ≥ 0.4 and pooled SBI ≥ 0.4. Binary maps were generated using threshold metrics maximizing sensitivity and specificity (Liu et al., 2016). TSS-based thresholds generally yield larger predicted ranges and tend to underestimate range losses relative to metrics like Matthew's Correlation Coefficient (MCC) or the F-measure, thus rendering our estimates conservative lower bounds for extinction risks due to climate and land use changes (Hellegers et al., 2025).

Prediction uncertainty was evaluated using 'extra_eval' from "flexsdm" version 1.3.3, employing the Shape metric (Velazco et al., 2023) to detect extrapolation beyond environmental conditions in training data. We applied a dynamic thresholding method based on taxa's extent of occurrence (EOO) percentiles (<12.5%, 12.5-50%, 50%-75%, >75%) to adapt extrapolation limits according to inferred niche breadth, following Velazco et al. (2023). Areas of high extrapolation uncertainty were removed from habitat suitability and binary maps. Non-zero cells in each taxon's clamping mask were set to NA to mitigate prediction artifacts (Elith et al., 2010).

Variable importance was assessed using 'ecospat.ESM.VarContrib' from "ecospat" 3.1, quantifying contribution via adjusted ratios comparing model weights with and without each variable. Ratios >1 indicate above-average influence.

Future range shifts were predicted with the 'BIOMOD_RangeSize' function from "biomod2" 4.2.4, assuming minimal dispersal for all orchids. Although species-specific dispersal would improve realism, such data remains difficult to estimate for multi-taxon studies. The analysis thus considers eight scenarios summarized in the following table:

**Table 1**. All projections generated in this paper. We modelled our species under an Ensemble of 15 GCMs for the baseline period and for two future periods, namely the 2050s and the 2070s, for two Representative Concentration Pathways (RCPs), namely the RCP 4.5 & 8.5 and two Shared Socioeconomic Pathways (SSPs), SSP1 and SSP5.

| Climate | Land use | Time-period | ID | Name |
|---|---|---|---|---|
| Mild | Mild | 2050 | 1 | en45_2050_ssp1 |
| | | 2070 | 2 | en45_2070_ssp1 |
| | Deep | 2050 | 3 | en45_2050_ssp5 |
| | | 2070 | 4 | en45_2070_ssp5 |
| Intense | Mild | 2050 | 5 | en85_2050_ssp1 |
| | | 2070 | 6 | en85_2070_ssp1 |
| | Deep | 2050 | 7 | en85_2050_ssp5 |
| | | 2070 | 8 | en85_2070_ssp5 |
| Current | | 2015 | 0 | current |

Note, for the SDMs used here, the analysis was restricted to the UK and is therefore likely to be subject to significant niche truncation. The current results were generated using an analytical pipeline, which could be made more current and robust and for which a revised analytical pipeline is being developed.

**2.4 Statistical analyses of trends and extinction**

Following an earlier paper (Charitonidou et al 2025) we define as *statistically significant* a trend whose probability of occurrence, under the assumption of natural change, is less than a specified level, of 5% . Given that the natural environment exhibits autocorrelation over all timescales, we need to control for the natural trends and cycles within the environment before we can decide whether a trend is indeed significant. The canonical stochastic process for autocorrelated variability is termed 1/*f*-noise (Halley, 1996), which describes well the fluctuations in real ecological populations and various proxies thereof (Inchausti and Halley, 2002). For each species in the dataset, we counted the number of occupied hectads for each of the four time-periods. This yields a proxy for the species' abundance as a function of time. For estimating the significance of the trend assuming 1/*f*-noise environmental variability, we used a Monte-Carlo method (Halley, 2009). We generated two contrasting stochastic models of environmental variability. 1/*f*-noise, also known as pink noise, was used to generate many simulated trajectories of annual occupancy to compare with the observed series. Values of the 1/*f*-process were generated for each year and sampled at each of the four time-periods was used to obtain the significance value. For the UK, this yielded a set of 14 species that were clearly declining, given a 1/*f*-noise environment. For this set we estimated the time to extinction by applying a trend analysis – by projecting forward in time to 2050 and 2070 and seeing at what point the fitted trend reaches zero.

### 3. Results
### 3.1 SDM-based projections of species richness and changes in 2050 and 2070

Following the methods described in the Section 2.1, we made projections for orchid species richness in the future, for the different scenarios described in Table 1. Fig. 1 shows the expected species richness in the different scenarios. Clearly there are substantial losses in all the different scenarios. However, greater losses (fewer darker reds) are observed at the 2070s compared to the 2050s. Also, the more intense and deeper scenarios (e.g. en85_2070_ssp5) lead to greater losses. These losses are not limited to the Great Britain. In Ireland, which shows increases of orchid biodiversity throughout the four time periods for the BSBI data, the prediction is also for there to be large losses. All scenarios were performed and yielded the maps shown in Fig1. It is clear that in several of the scenarios, the occupancy is decreased substantially in the future. We also note that there is little evidence shown of any northward "climatic" shift in these predictions. It is notable that the species richness in Scotland is lower in all scenarios. This is in marked contrast to the results of the earlier paper (Charitonidou et al 2025) in which a Northward trend was visible in the UK data.

### 3.2 Predictions of loss and gain based on distribution projections

Figure 2 shows that the SDM projections foresee substantial losses of occupancy in the UK. The SDM projections show losses for most of the species. A notable feature is the increases expected in Wales and Western England. Also for southern England (esp. Cornwall) there are increases in the 2050s, though not much in the 2070s. For all scenarios, there are losses expected in Scotland, in contrast to what might be expected on the basis of climate change. Also, few increases are seen in Ireland, it is mostly losses.

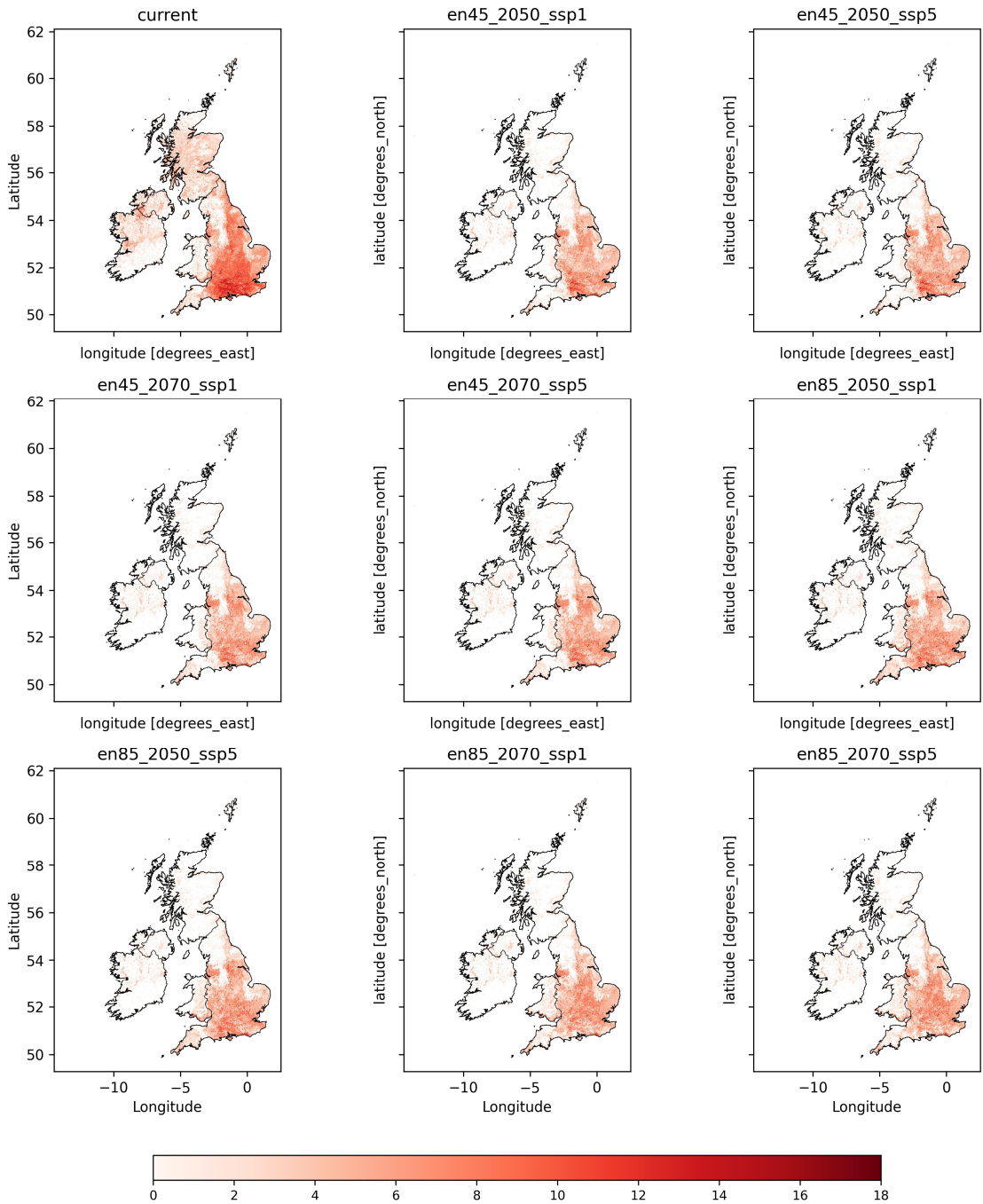

**Fig.1**. Projections for orchid species richness of the British Isles for eight different scenarios described in Table 1. Deeper shades of red are higher Species richness. Note that the more intense and deeper scenarios (e.g. en85_2070_ssp5) lead to greater losses. Maps show the spatial distribution of total predicted biodiversity for each scenario, at the monad resolution, as well as areas of absolute gain and loss of species number compared to the predicted current baseline, highlighting the spatial patterns.

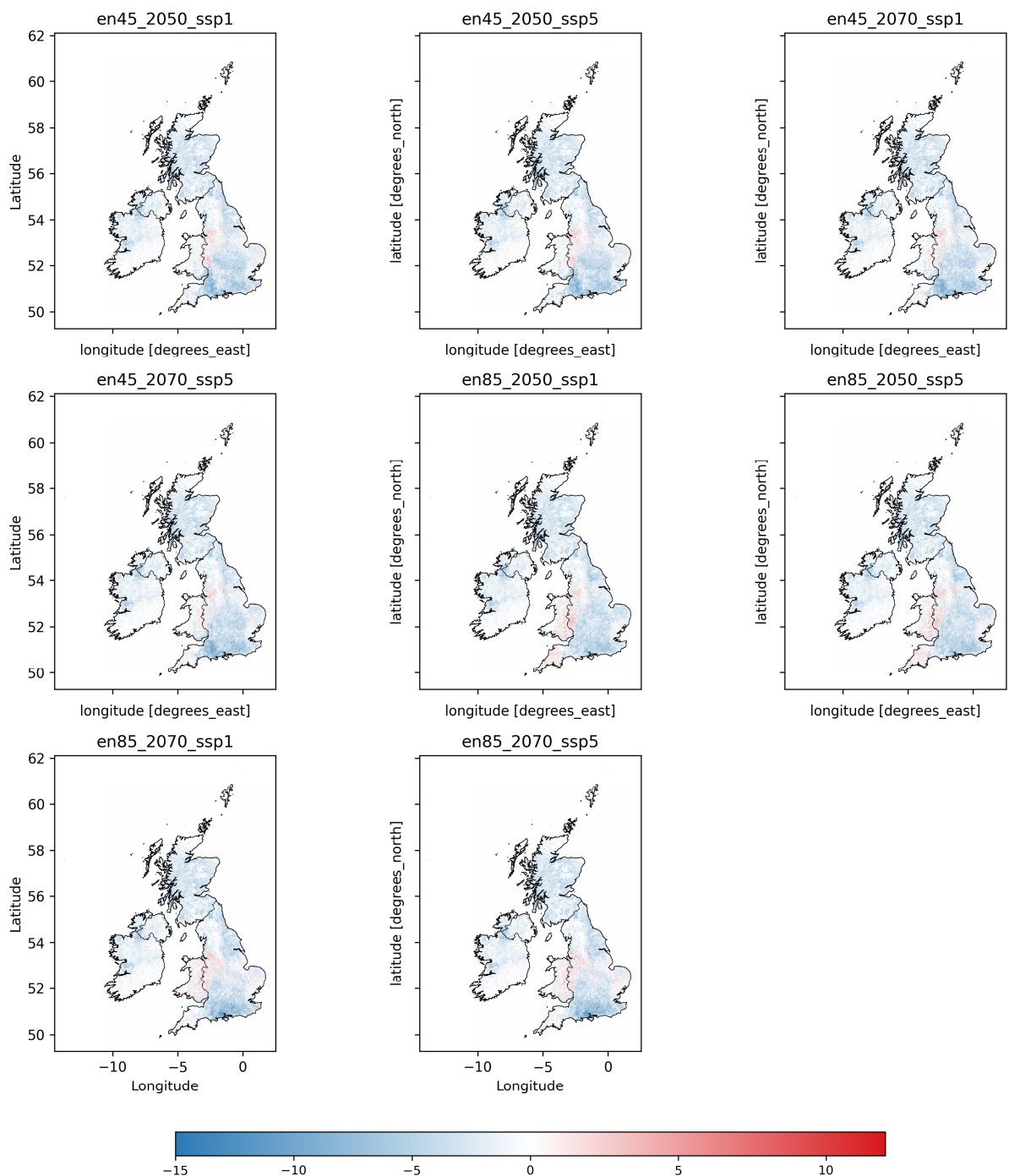

**Fig.2**. Projections for changes in orchid species richness for eight different scenarios described in Table 1, showing overall losses (blue) and gains (red) for each species as percentage change from current coverage. Percentage change in predicted suitable area relative to the predicted current conditions for each species Note that the more intense and deeper scenarios (e.g. en85_2070_ssp5) lead to greater losses. Maps show the spatial distribution of total predicted biodiversity for each scenario, as well as areas of absolute gain and loss of species number compared to the predicted current baseline, highlighting the spatial patterns.

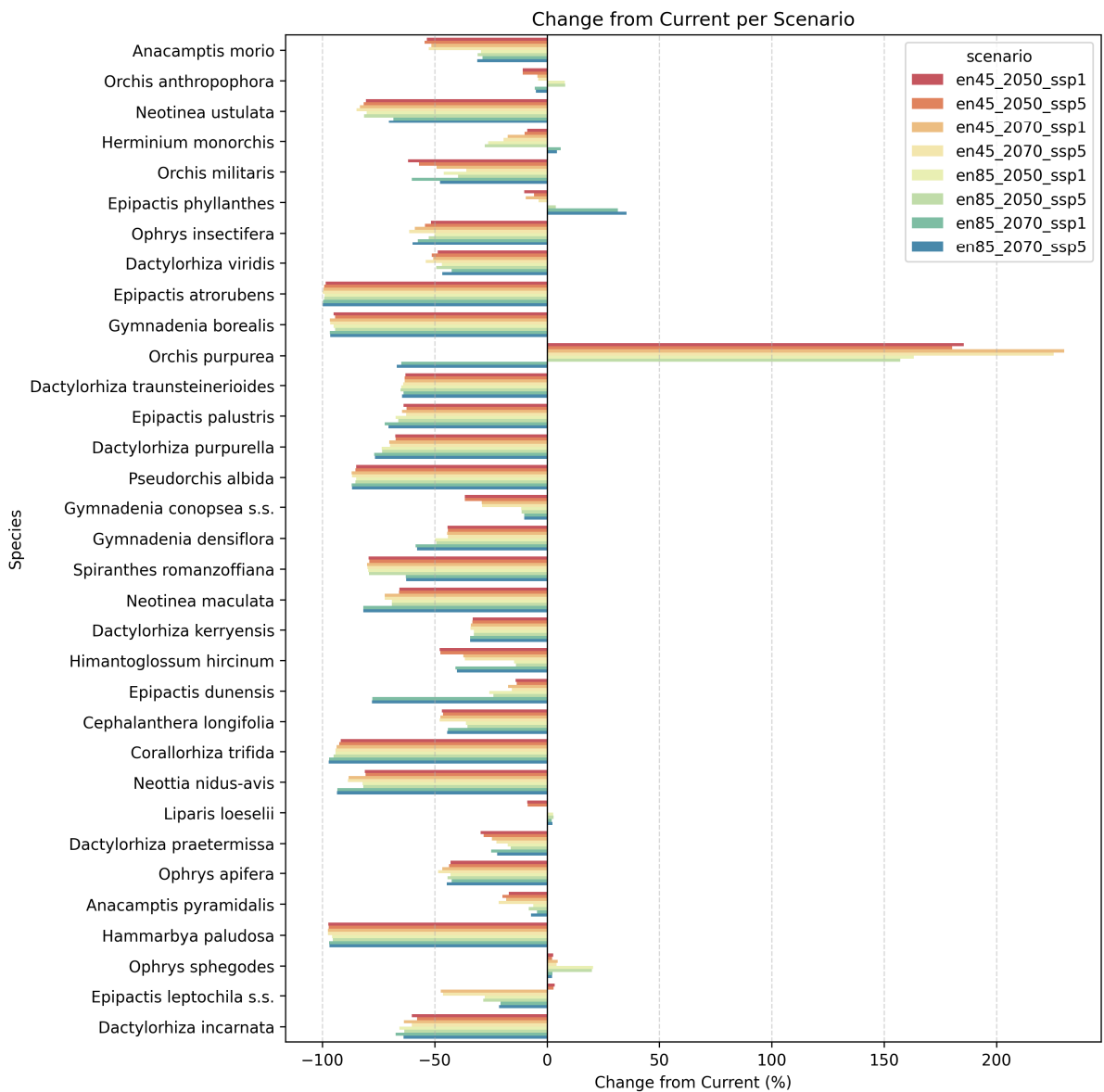

**Fig.3**. Bar chart showing overall losses and gains for each species as percentage change from current coverage. Percentage change in predicted suitable area relative to the predicted current conditions for each species under two emission pathways (EN45 and EN85) and two socioeconomic scenarios (SSP1 and SSP5), shown for the years 2050 and 2070. Each panel represents a unique combination of climate model and socioeconomic scenario as in Table 1. Bars indicate the direction and magnitude of change, with red denoting area loss and green denoting area gain relative to current distributions.

According to, the predictions of our SDM framework in Fig 3, most species' distributions will contract. For most species the number of monads occupied will decline. This will be especially true for *Gymnadenia borealis*, *Hammarbya paludosa*, *Corallorhiza trifida* and *Epipactis atrorubens*, which will all disappear from over 90% of their locations. Only three species will increase their occupancy: *Orchis purpurea*, which will increase by over 160%, and *Ophrys sphegodes* and *Liparis loeselii*, who will increase marginally.

Page 11

**3.3 Predictions of extinction or growth based on significant trends**

For the UK, we took the set of 14 species that were clearly declining, given a 1/*f*-noise environment. For this set we estimated the time to extinction by applying a trend analysis – by projecting forward in time to 2050 and 2070 and seeing at what point the fitted trend reaches zero. The results are found in Table 2.

**Table 2**. Species that were clearly declining (at the 10% significance level), given a 1/*f*-noise environment, following (Charitonidou et al., 2025) and predictions of extinction on this basis. The columns correspond to Species, Average occupancy in hectads over all 4 time-periods, the loss rate according to the fitted trend, the *p*-value according to Monte-Carlo analysis, the estimated time to extinction, from 2025, and whether the species is expected to be extinct by 2050 or 2070.

| Species | Abundance (Hec.Av) | Loss rate (Hec./dec.) | *p* | Time to extinction | Extinct 2050 | Extinct 2070 |
|---|---|---|---|---|---|---|
| *Ophrys insectifera* | 127.5 | 15.1 | 0.001 | 50 | | |
| *Neotinea ustulata* | 56.0 | 9.3 | 0.008 | 26 | | X |
| *Herminium monorchis* | 33.0 | 5.4 | 0.010 | 27 | | X |
| *Platanthera bifolia* | 393.8 | 46.5 | 0.021 | 50 | | |
| *Liparis loeselii* | 8.5 | 1.4 | 0.031 | 27 | | X |
| *Cephalanthera rubra* | 3.5 | 0.3 | 0.042 | 71 | | |
| *Neotinea maculata* | 0.3 | 0.2 | 0.043 | 0 | X | X |
| *Spiranthes spiralis* | 300.8 | 21.5 | 0.049 | 105 | | |
| *Dactylorhiza viridis* | 379.8 | 41.5 | 0.055 | 57 | | |
| *Neottia nidus-avis* | 349.5 | 30.4 | 0.079 | 80 | | |
| *Orchis purpurea* | 18.0 | 1.5 | 0.083 | 84 | | |
| *Pseudorchis albida* | 130.8 | 14.0 | 0.084 | 59 | | |
| *Epipactis leptochila* | 24.3 | 3.2 | 0.096 | 40 | | X |

According to this analysis, of the species analysed only one goes extinct by 2050, while five are extinct by 2070. By contrast, *Neotinea maculata* is identified as "already extinct because it has not been found in the last 3 time periods. Species that are expected to be extinct by 2070 are those with strong rates of decline that are already at relatively low densities, namely *Neotinea ustulate*, *Liparis loeselii*, *Herminium monorchis* and *Epipactis leptochila*. In the table above, both *Ophrys insectifera* and *Dactylorhiza v*iridis have strong downward trends, but there is no expected danger of extinction by 2070 because the abundance is relatively higher.

**Table 3**. Species that were clearly growing (at the 10% significance level), given a 1/*f*-noise environment, following (Charitonidou et al., 2025). The columns correspond to Species, Average occupancy in hectads over all 4 time-periods, the growth rate according to the fitted trend, the *p*-value according to Monte-Carlo analysis.

| Species | Abundance (Hec.Av) | Growth rate (Hec./dec.) | *p* |
|---|---|---|---|
| *Dactylorhiza traunsteinerioides* | 34.8 | 4.6 | 0.01 |
| *Gymnadenia densiflora* | 90.8 | 7.5 | 0.03 |
| *Epipactis dunensis* | 26.5 | 4.1 | 0.03 |
| *Epipactis phyllanthes* | 88.8 | 3.1 | 0.03 |
| *Ophrys apifera* | 814.0 | 58.9 | 0.05 |
| *Spiranthes romanzoffiana* | 16.8 | 1.3 | 0.06 |
| *Serapias lingua* | 1.0 | 0.3 | 0.06 |
| *Corallorhiza trifida* | 53.8 | 4.1 | 0.06 |
| *Anacamptis pyramidalis* | 709.0 | 45.2 | 0.07 |

## 4 Discussion

Concerns about the populations of orchids are widespread and several scientific studies have argued that orchids are "in decline" in Britain and Europe. However, definitive answers remain elusive and our earlier paper study (Charitonidou et al., 2025) for both Great Britain and Ireland over the 90-year period covered by the BSBI atlases, found a dominant pattern of decline in orchid biodiversity, and most of the significant trends were decreasing in Great Britain. However, the pattern was the opposite in Ireland, with most significant trends positive and several high-diversity hotspots appearing in recent years. In making predictions from the future based on such studies, one must project these trends into the future.

When we use these trends to predict the state of occupancy in 2070 we see that only one is extinct by 2050, while five are extinct by 2070. Species that are expected to be extinct by 2070 are those with strong rates of decline that are already at relatively low densities, several species have strong downward trends but are expected to persist to 2070 because their abundance is currently high. Thus, on the basis of observed trends in the BSBI database, relatively little is expected to change, even by 2070. This does not take into account changes in land use and accelerating levels of climatic change and it is possible that these will significantly alter the fate of orchid species in this time frame.

The use of SDMs provides an alternative approach to this problem. In this case we can incorporate the role of changes in land use and different levels of climatic change by testing different scenarios. Our study here also used monads rather than hectads, enabling us to make predictions at a higher resolution. Our study made projections for orchid species richness for eight different scenarios (Table 1). According to, the predictions of our SDM framework in Fig 3, most species' distributions will contract. Here we see there are substantial losses in all the different scenarios, greater losses are observed in the 2070s compared to the 2050s. Also, as expected, the more intense scenarios of climate change and heavier land use lead to greater losses. These differ considerably from predictions based on current trends in Table 2 and they are not limited to the Great Britain but are also expected in Ireland. Ireland enjoys increases of orchid biodiversity throughout BSBI data, but in our SDM predictions this is reversed. Indeed, significant losses are predicted even by 2050. The predictions show no northward "climatic" shift but a increases of orchid diversity in the South and West of Great Britain and a decrease in the South of England and in Scotland. This is in marked contrast to the current trends seen in our earlier paper (Charitonidou et al 2025) in which a Northward trend was visible in the UK and Soutwest in Ireland.

In the SDMs used here, the analysis was restricted to the UK. Since many of the species are present widely in Europe and beyond, it is likely that these are subject to significant niche truncation. This may explain some of the more severe declines we see in the predictions for range contraction. An analysis using wider geographical data and also incorporating some improvements to the analytical pipeline in terms of robustness and performance, is the subject of a manuscript in preparation. If we use wider maps for niches and abiotic conditions for each species, then we probably will have less niche truncation and greater robustness in the face of changing conditions. This will lead to greater convergence between SDM based methods and more traditional forecasting methods based on orchid population dynamics. It will also give us a more coherent insight into the future distribution of orchids in Great Britain and Ireland.

**Author Contributions**
**SM**: Ran software. Contributed to writing and editing, formal analysis, data acquisition & curation. **KK**: Developed SDM Methodology and wrote software. Contributed to running software, data acquisition & curation, editing and funding acquisition. **MC**: Contributed to editing, data acquisition and curation, funding acquisition and project administration. **JMH** wrote the paper. Contributed to editing, formal analysis, funding acquisition and project administration.

**Funding**
The author(s) declare financial support was received for the research and/or publication of this article. The research project was supported by the Hellenic Foundation for Research and Innovation (H.F.R.I.) under the "2nd Call for H.F.R.I. Research Projects to support Faculty Members & Researchers" (Project Number: 3972).